%% file: main.tex
\documentclass[10pt,letterpaper, twocolumn]{article}

\usepackage[utf8x]{inputenc} 
\usepackage{fullpage}
\usepackage{setspace}
\usepackage[usenames]{color}
\usepackage{url} 
\usepackage{mdwlist} 
\usepackage{hyperref}
\usepackage[sf,bf]{caption}
\usepackage[bf,compact]{titlesec}
\usepackage{graphicx}
\usepackage{multirow}

\usepackage[T1]{fontenc}
\usepackage{times}
\usepackage[scaled=0.9]{helvet}

\begin{document}

\title{Building Better Incentives for Robustness in BitTorrent}
\author{
Seth James Nielson, Caleb E. Spare, Dan S. Wallach \\
{\small seth@sethnielson.com, cespare@rice.edu, dwallach@cs.rice.edu} }

\date{Computer Science Department, Rice University}
\maketitle

\begin{abstract}
\input abstract
\end{abstract}

\section{Introduction}
\label{intro}
\input{intro}

\section{Background}
\label{background}
\input{background}

\section{Incentives Design}
\label{incentivesdesign}
\input{incentivesdesign}

\section{Methodology}
\label{methodology}
\input{methodology}

\section{Evaluation}
\label{evaluation}
\input{evaluation}

\section{Discussion and Future Work}
\label{discussion}
\input{discussion}

\section{Related Work}
\label{related}
\input{related}

\section{Conclusion}
\label{conclusion}
\input{conclusion}

\section*{Acknowledgements}
The authors wish to thank Johan Pouwelse for collecting and sharing his traces
from many real BitTorrent swarms.  We also acknowledge Ed Knightly, Eugene Ng, Dan Sandler, and Devika Subramanian for many helpful discussions on this paper.  Scott Crosby offered incredible assistance in performance tuning our simulator.  This research was supported, in part, by NSF grants CNS-0524211 and CNS-0509297.

{\small
\bibliographystyle{abbrv} \bibliography{peer2peer,proposal,twngan,attackstaxonomy,prior_work}
}
\end{document}

%% file: abstract.tex
BitTorrent is a widely-deployed, peer-to-peer file transfer protocol
engineered with a ``tit for tat'' mechanism that encourages
cooperation.  Unfortunately, there is little incentive for nodes to
altruistically provide service to their peers after they finish
downloading a file, and what altruism there is can be exploited by
aggressive clients like BitTyrant.  This altruism, called seeding, is
always beneficial and sometimes essential to BitTorrent's real-world
performance.  We propose a new long-term incentives mechanism in
BitTorrent to encourage peers to seed and we evaluate its effectiveness
via simulation. We show that when nodes running our algorithm reward 
one another for good behavior in previous swarms, they experience as 
much as a 50\% improvement in download times over unrewarded 
nodes. Even when aggressive clients, such as BitTyrant, participate in the 
swarm, our rewarded nodes still outperform them, although by smaller
margins.

%% file: intro.tex
Peer-to-peer file transfer protocols provide scalable architectures
for distributing large files.  The core idea is to have peers
participating in the download also contribute upload service back to
the system, thus scaling the available bandwidth as more peers
join. Even centralized services with large network connections can be
overwhelmed by flash crowds, while p2p services can ostensibly
continue to scale, even in such extreme scenarios.

In the practical world, however, scalability and stability in p2p
systems are limited
by the cooperation of the participants. These
systems only have as much bandwidth as is collectively donated.
Proper behavior cannot necessarily
be enforced; participants are going to behave \emph{rationally},
taking whatever steps maximize their own benefit without particularly
caring about the well-being of other peers.
Consequently, the default behavior of most participants is to consume
and not contribute. This is often called the ``free rider'' problem.

BitTorrent~\cite{bittorrent-incentives} mitigates the free rider problem by rewarding uploads by
granting faster downloads through a ``tit for tat'' (TFT) protocol,
thus making cooperation a rational behavior.  This design
has been highly successful, enabling BitTorrent's wide acceptance in
the Internet community. While there is no consensus on the true
amount of BitTorrent data in-flight today, it is clear that the number is
large at somewhere between one-third and one-half of all Internet 
traffic~\cite{bittorrent_traffic_original, torrentfreak_myth, wired_bittorrent_traffic,
ipoque_bittorrent_traffic}. 

Despite the practical success of BitTorrent, numerous researchers have
exposed weaknesses to the TFT incentives
mechanism~\cite{bittyrant1_2007,tian,sirivianos,liogkas}. One prominent 
weakness is the significant level of altruism that remains in the
system despite the TFT mechanism. More specifically, many peers still
contribute significant upload bandwidth without necessarily improving
their download performance. Such contributions are produced by
asymmetries in upload and download bandwidth as well as by altruistic
BitTorrent behaviors like seeding and optimistic unchoking. (Section~\ref{sec:background:ambient}
discusses this ``ambient altruism'' in detail.)

These exploits are not simply theoretical. BitTyrant~\cite{bittyrant1_2007} 
takes advantage of the intrinsic altruism to achieve high download rates
while reducing upload contributions. Most BitTorrent clients can be easily configured to rely
exclusively on leeching, and some research suggests this is effective
despite the TFT incentives~\cite{bitthief,sirivianos}.

Our goal in this work is to reduce the altruism in BitTorrent seeding by adding
incentives to the seeding component of the protocol. We present the design and
evaluation of our seeding reward algorithm which requires a minor change to BitTorrent
in the form of a long-term identifier for participating clients. Through
simulation we demonstrate that rewarded
peers get better performance than unrewarded peers. This differential creates
an incentive for rational nodes to switch into the rewarded population. We further
 show that the rewarding mechanism improves node
performance even when some portion of the swarm is composed of BitTyrant nodes.

In the remainder of the paper, we first review the operations and altruism of BitTorrent in Section~\ref{background}
as well as an overview of the BitTyrant variant. 
Sections ~\ref{incentivesdesign} and~\ref{methodology} present
our algorithm and the methodology we use to evaluate 
its performance. Our results are detailed in Section~\ref{evaluation} and further analyzed in Section~\ref{discussion}.
We close with a discussion of related work in 
Section~\ref{related} and our conclusions in Section~\ref{conclusion}.

%% file: background.tex
BitTorrent~\cite{bittorrent-incentives} is a highly successful and 
popular peer-to-peer protocol which aims to enable efficient, 
rapid distribution of potentially large amounts of data to a group 
of clients. It is designed to utilize the available upload bandwidth 
of the clients to scale the capacity of the system to support many 
users and has built-in mechanisms to incentivize participation 
in this scheme.

\subsection{The BitTorrent Protocol}
A \emph{torrent} is a file or a set of files users wish to download. The data is divided into equal-sized \emph{pieces}, typically 256KB, which are further subdivided into small \emph{blocks}. A central node called the \emph{tracker} keeps track of the peers participating in 
the distribution of a torrent.
The tracker does not serve the actual content, but instead serves as a 
rendezvous point for peers to discover
one another.

BitTorrent clients use a file of metadata, called a \emph{torrent file}, 
to begin downloading content. This file, typically downloaded from a 
traditional web server, specifies the address for the tracker as well 
as information about the files to be downloaded, including names, 
sizes, and SHA-1 checksums for each piece.

The set of clients working on downloading a given torrent is referred to as a \emph{swarm}.  Clients notify the tracker as they join and leave the swarm, as well as every 30 minutes they are active within the swarm.  To discover other clients, a client may query the tracker, which gives it a random subset of the active peers.  (A variety of extensions exist which supplement the tracker, including a gossip protocol as well two DHT-based schemes.)  Once it has a set of peers, a client establishes TCP connections to its peers, forming a \emph{neighborhood} with whom it shares information about which pieces it has and has not completed downloading.  A legitimate publisher might establish one or more official \emph{seeds}, which provide round-robbin, best-effort service to anyone who asks.  These seeds are then supplemented by altruistic peers who seed after they finish their downloads.


\subsection{BitTorrent Strategies}
Popular BitTorrent clients employ a number of strategies to encourage 
fair participation in uploading and to deal with a variety of 
corner cases~\cite{bittorrent-incentives}.

A client only uploads to a small number of peers in its neighborhood at any 
given time. This group of nodes is called the client's \emph{active set}. The size of
the active set is typically four, although both the reference implementation and
BitTyrant~\cite{bittyrant1_2007} note that this number should scale with maximum upload
bandwidth capacity. The majority of the nodes in the active set are the nodes that
have given the best service over a rolling 20 second average. The client saves one
or two slots in the active set for the exploration of new neighbors.
\emph{Optimistic unchokes} pick a random peer
every 30 seconds, allowing the client to search for better neighbors while
also bootstrapping newly joined clients that have not yet downloaded anything to
share.

BitTorrent clients share current status information with other clients to indicate which pieces are completely downloaded. Clients will bias their block requests to complete one piece before they begin downloading a different piece.
To pick a piece to download, BitTorrent follows a \emph{rarest first} policy, where a client picks pieces based on lowest availability within its neighborhood. The exception to this rule is for new clients, which need a complete piece before they can advertise any content for upload. In this case, they instead pick a random piece.

When a block has been requested, a client does not reissue the request until either the block is received or the request times out. This can be a problem when a user has received most of the pieces in a file and has just has a few outstanding requests to go.  If the final peers are slow or unresponsive, the system might never finish. In this case, the client goes into \emph{endgame mode} and sends redundant requests for any missing blocks to its peers; as they are received the client sends messages to the remaining peers to cancel unnecessary requests.

\subsection{Ambient Altruism and BitTyrant}
\label{sec:background:ambient}
BitTorrent aims to reduce the free-rider problem, but it is not
intended to eliminate altruism in the system.  Instead, BitTorrent
aims to ensure that a node will experience significantly improved
performance if it participates in TFT trading, rather than leeching.
Consequently, altruistic features remain in the protocol and pose two
separate, but related, problems. First, a client can reduce or
eliminate its own altruistic participation, reducing the overall swarm
performance. Second, if a client can recognize peers that are
participating altruistically, it may be able to obtain sufficient
service from these peers to find it unnecessary to deal with those
that require cooperation.

Two significant sources of altruistic contributions are seeding and
optimistic unchoking. Seeding is inherently altruistic under the
current BitTorrent protocol. The altruism of optimistic unchoking is
more complex. The optimistic unchoke operation
is BitTorrent's method of searching the peer space for better TFT
service.  An unchoke that results in improved service because a better
peer is found is clearly not altruistic, but unchokes are performed with
random peers, rather than being biased away from known leeches.  This
means that BitTorrent's standard unchoking behavior can still provide
a source of altruism, to the benefit of leeches.

Differences between peer bandwidth capacities also produce altruism.
When a normal BitTorrent client unchokes a peer, it sends data as fast
as the TCP stack will go, so peers with faster network connections
will tend to give more out than they get in return when dealing with
slower peers.  Of course, two fast peers with content to trade will be
more likely to establish TFT trading with one another than a fast peer
and a slow peer.


BitTyrant is a strategic BitTorrent variant that exploits ambient
altruism and reduces its own altruistic
contributions~\cite{bittyrant1_2007}. BitTyrant was designed to
download as fast as possible while contributing the minimum amount
required to achieve it. To achieve this, BitTyrant abandons
BitTorrent's policy of giving each member of the active set an equal
share of its upload bandwidth. Instead, BitTyrant unchokes as many
neighbors as possible but limits the speed of each upload stream to be
only as much as is necessary to obtain reciprocation.

This scheme does not work for other BitTyrant nodes, however, and two BitTyrant nodes
must enter a special mode when dealing with each other. In 
Section~\ref{section:eval-bittyrant}, we will describe this special mode in detail
and demonstrate how it can be used as part of a defense against BitTyrant's behavior.

%% file: incentivesdesign.tex
Our incentives design for seeding in BitTorrent requires that the BitTorrent
protocol support some form of long-term identifier. The basic concept for
our algorithm is that BitTorrent clients recognize seeders from previous
swarms and this is impossible without these IDs. Fortunately,
the exchange of long-term identifiers can be built into the peer handshaking process
in a backwards compatible fashion. Clients without a long-term ID
are simply assumed to have no history. It is also worth noting that some 
clients~\cite{tribler} already support an optional long-term ID.

Our proposed design consists of an observation phase and a reward phase. The
observation phase is in effect whenever the node is receiving seeding bytes,
or bytes received from a neighboring peer without the expectation of 
TFT reciprocation.
The detection of seeding bytes, in our basic implementation, is based on
first-hand, verifiable information only.
Obviously, it is possible that the neighbor is only pretending to seed, but
from the observing node's perspective, all bytes received without giving any
bytes in return are seeded bytes.

The reward phase occurs when the node is in seeding mode. The goal is to schedule
outbound seeding with higher priority given to peers who have seeded in the
past. To do this, the algorithm first computes a score for each node; nodes
who seeded get higher scores.  These scores are used to initialize a scheduler,
giving more slots to nodes with higher scores.  While virtually any scheduling
algorithm would suffice, we chose to use lottery scheduling~\cite{lotteryscheduling}.  Each peer gets at least one ticket, but peers that seed get additional tickets in proportion to the logarithm of the number of bytes we have received from them in seeding.

Obviously, a node that chooses to be a good citizen and seed
may not be rewarded at all in the future. For node $A$ to be rewarded
by node $B$, $A$ must seed to $B$ and then $B$ must seed
to $A$ in some subsequent swarm. That means that both nodes must
interact repeatedly over time. For any real benefit to the algorithm, a
group of nodes must interact repeatedly.

We note that a Sybil attack~\cite{sybil} is possible against
this protocol.  For example, malicious nodes could create a large
number of false identifiers, gaining additional shares of the
bandwidth.  We deal with this by reserving a percentage of a seeder's
upstream bandwidth for other known seeders.  Sybil attackers may well
fight it out for the remaining unreserved bandwidth, but there is a
larger pool of bandwidth available if they cooperate.

Another possible Sybil attack would be a \textit{reincarnation attack}~\cite{Nielson05Taxonomy} where a client sheds an old identifier
for a new identifier in every swarm to erase previously observed bad behavior.
Such behavior would be unhelpful to the node, however, because a fresh identifier
begins with no rewards at all.  Rewards only come with observed 
good behavior.



%% file: methodology.tex
\subsection{Simulator}
We chose simulation as our primary method for analyzing incentives and
altruism in BitTorrent. The advantages of a simulator over real world tests
or the use of network emulation lies primarily in the repeatability
of the experiment and the time required to run the experiment. Our research
requires comparison of algorithms against one another as well as 
experimentation with hundreds of combinations of parameters. Repeatability
and fast time to completion were both incredibly helpful.

Several BitTorrent simulators exist but they did not fully meet our needs. One
simulator from MSR~\cite{msr_simulator_2005} does not implement asynchronous communication
nor does it capture some BitTorrent details, such as piece chunk transmission,
that we deemed necessary. An ns-2~\cite{egeretal_upgrade_07} BitTorrent simulator was also
available, but it simulates TCP effects and other network level details that were
too low level for our purposes. GPS~\cite{gps_2005} is a general purpose p2p simulator
that includes a BitTorrent module and simulates at about the same level of granularity as our work.
GPS is written in Java and our work appears to run faster. 

To meet our objective, we have designed an optimized C++ simulator with a
Python front end for simulation setup and execution. Our simulator allows
swarms of thousands of clients, with several hundred running simultaneously,
many times faster than real-time. To illustrate this, we ran a series of
tests on an Athlon 2.4Ghz dual-processor server with 4GB
of RAM and running with the Linux 2.6.9 kernel. These tests employed a
 simple swarm where 
a given number of clients arrive simultaneously and join the swarm.
There is only a single seed for the swarm. We fix
the file size at 100MB, the seed's upload capacity at 512Kbps, 
and each client's bandwidth at 56Kbps, symmetric for uploads and downloads. The results for
various swarm sizes is shown in Table~\ref{table:benchmark1}.
These results show that the time required to simulate the swarm
is proportional to the number of peers.

\begin{table*}
\centering
\begin{tabular}{ | r | r | r | r |  r |}
\hline
$n$ & Sim Time (hours) & Real Time (hours) & Messages & Memory (MB) \\
\hline 
10 & 5.86 & 0.004 & 233,950 & 20 \\
\hline 
100 & 4.77 & 0.07\hspace{0.5em} & 1,381,715 & 60 \\
\hline 
1000 & 5.24 & 0.86\hspace{0.5em} & 13,635,955 & 492  \\
\hline
\end{tabular}
\caption{Basic simulator performance as the number of simulated nodes ($n$) grows.\label{table:benchmark1}}
\end{table*}

\subsection{Simulation Setup}
\label{sec:methodology:setup}
All the evaluations in this paper are based on a flash-crowd, 1GB file BitTorrent swarm.
We used a total population of 2000 
DSL clients with a range of download bandwidths from 128Kbps to 5Mbps. Each client's
upload bandwidth is precisely half of its download bandwidth. 
To obtain reasonable churn, we make use of real-world BitTorrent traces taken in 2005 by
Johan Pouwelse. These traces provide realistic join
times for flash-crowd behavior in real swarms.

\begin{figure}
\centering
\includegraphics[scale=.3]{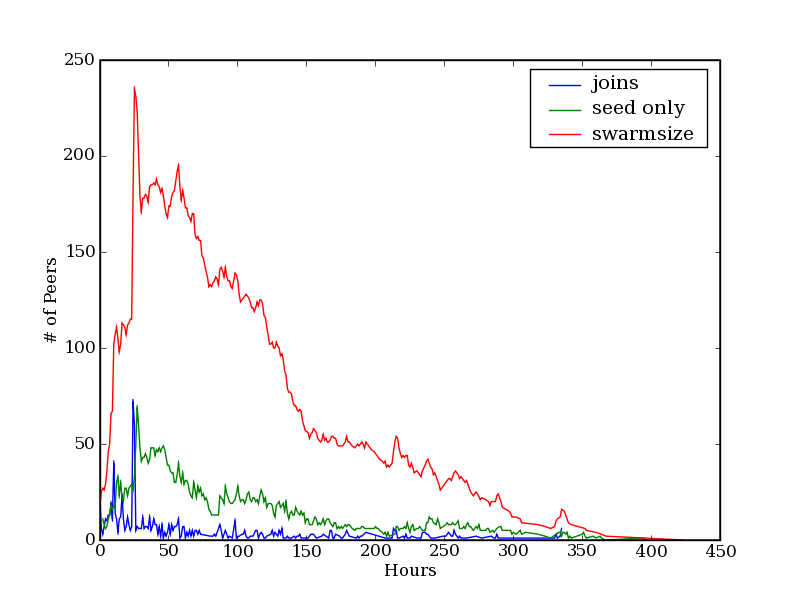}
\caption{Simulated swarm membership over time based on a real-world trace
from a flash-crowd swarm.\label{figure:churn}}
\end{figure}

Each simulation is also configured with experiment-specific parameters. The significant
parameters are:

\paragraph{Seeding Time} The 2000 clients of the swarm are assigned one of three
seeding population types. \textit{Altruistic} clients will seed for 24 to 48 hours
after their download is complete. \textit{Standard} clients seed for one to two hours. \textit{Leech} clients terminate their connection
immediately after downloading the object. These values are based on 
why peers choose to seed; altruistic clients intentionally stay around to be helpful, standard
clients will continue running until the user notices the download is done and kills the client, and leech clients leave as quickly as possible. 
Even though these numbers are guesses, we have validated that a
swarm with 10\% altruistic nodes and 70\% standard nodes yields seed-to-swarm ratios similar
to those observed in a prior measurement study.
(Figure~\ref{figure:churn} in this paper closely resembles
Figure~5 in Pouwelse~et~al.~\cite{bt_measure2_2005}.)

\paragraph{Seeding Algorithm} Populations in the swarm can be assigned to use
different seeding algorithms. The standard seeding algorithm simply seeds round-robbin to all of the peers in a seed's neighborhood. We also support an ``incentives seeding'' algorithm, as described in Section~\ref{incentivesdesign}.

\paragraph{Incentives Seeding Parameters} For peers using the incentives
seeding algorithm, we can vary the bandwidth reservation for rewards as
a percentage of the total bandwidth; all incentives seeding nodes will use the same reservation percentage in a given simulation run. Also, for nodes using our rewarding seeding
algorithm, we invent a past history for each one, assigning them a number of bytes
that they have seeded in the past. We similarly vary what portion of the population 
are aware of this history, allowing us to simulate everything from oracular
knowledge of every node's past behavior down to fragmentary knowledge that would be a more realistic approximation of prior, first-hand
observations. 

While oracular knowledge is unrealistic in practice, it allows us to
place an upper bound on the benefits of seeding policies that use this knowledge. First hand information is
more limited in scope but much more difficult to exploit~\cite{Nielson05Taxonomy}.
In our research we are assuming that there are no disjoint cliques of overlapping peers.  This would seem to adequately capture common classes of real-world behavior as we might expect from people who download related content, such as new episodes of TV shows released on a weekly basis.

\paragraph{Trading Algorithm} We have implemented both the regular BitTorrent TFT 
and the BitTyrant trading algorithms
in our simulator. Trading and seeding algorithms may be assigned
independently; a peer can use the BitTyrant trading algorithm and our incentives
seeding algorithm if desired. 

\subsection{Incentives Evaluation}
Our goal is to create an incentive for participants
in BitTorrent to seed. We will evaluate the effectiveness
of our algorithm by demonstrating that rewarded populations perform better
than unrewarded populations in our simulated swarms. By running the 
experiments under a variety of configuration parameters, we will characterize
how these parameters affect the success of our incentives algorithm.

In evaluating the performance of a node, our basic measurement
is the download efficiency, defined as the utilization of the peer's download pipe over its lifetime in the swarm.
Efficiency is a direct measure of the node's happiness, and it
is perfectly normalized. Any node, regardless of speed, cannot be happier than
when it has 100\% download utilization. 

\begin{table*}
\centering
\begin{tabular}{ | c | c | c | c | c | c |}
\hline
\multicolumn{3}{|c|}{Median Efficiency (\%)} & \multicolumn{3}{c|}{Median Download time (s)} \\
\hline
Altruistic & Standard & Leech & Altruistic & Standard & Leech \\
\hline
98.8 & 48.9 & 90.1 & 3304 & 7443 & 4402 \\
\hline
\end{tabular}
\caption{Comparison of median efficiency 
and median download time for the same experiment. 
\label{table:median_compare}}
%
\vskip 1em
\centering
\begin{tabular}{|c|c|c|r|}
\hline
Population & Average & Std. Dev & 95\% Confidence Interval \\
\hline
Altruistic & 98.0\% & 1.8\% & 4.1\% \\
\hline
Standard & 57.9\% & 8.5\% & 15.3\% \\
\hline
Leech & 87.6\% & 4.8\% & 8.2\% \\
\hline
\end{tabular}
\caption{Median efficiency, averaged over twenty different experimental runs,
differing only in the random seed.
\label{table:exp1_stats}}
%
%
\vskip 1em
\centering
\begin{tabular}{|c|c|c|r|}
\hline
Population & Average & Std. Dev & 95\% Confidence Interval \\
\hline
Altruistic & 97.9\% & 1.9\% & 4.6\% \\
\hline
Standard & 71.1\% & 7.9\% & 11.9\% \\
\hline
``Leech'' & 71.2\% & 7.6\% & 12.9\% \\
\hline
\end{tabular}
\caption{Median efficiency, averaged over twenty experimental runs as above, with the leech nodes replaced by standard nodes.
\label{table:exp2_stats}}
\end{table*}

Computing the efficiency $e$ is straightforward.
Let $k$ be the maximum download capacity of the node measured in bits per second (bps). Then
let $t_0$ be the time the peer connected to the swarm and let $t_d$ be the time that
it finished the download, where both values are measured in seconds. Finally,
let $n$ be the number of bits in the download object. Then
\[e = \frac{n/(t_d-t_0)}{k}.\]

\begin{figure}
\centering
\includegraphics[scale=.3]{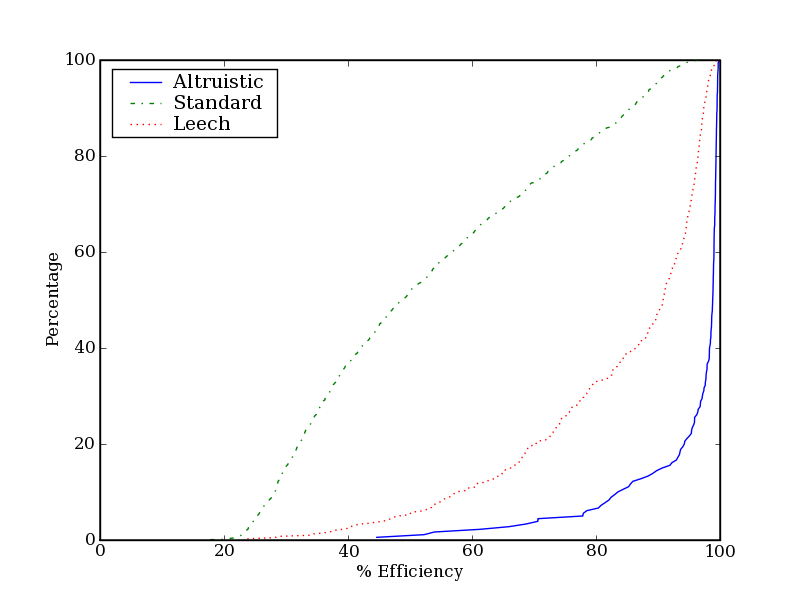}
\caption{Cumulative distribution of efficiency (bandwidth
utilization) over different populations in the same swarm.
\label{figure:sample_efficiency}}
\centering
\includegraphics[scale=.3]{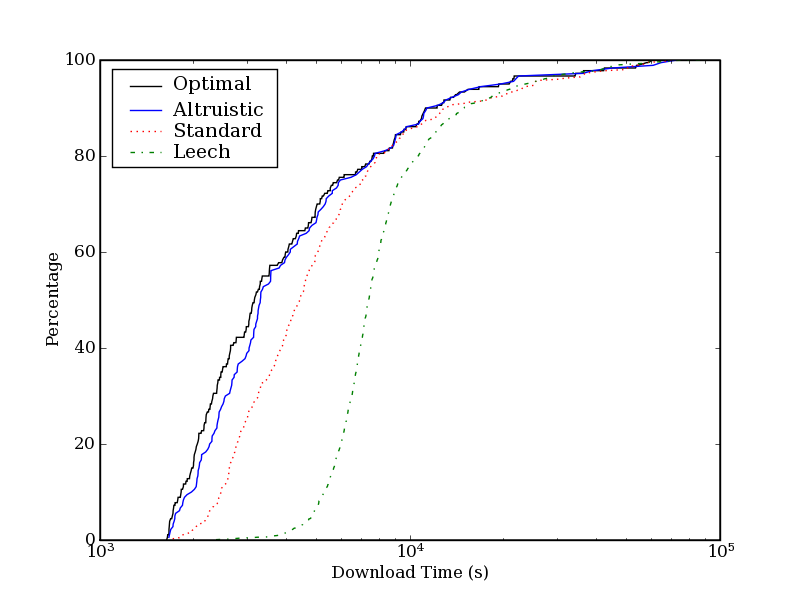}
\caption{Cumulative distribution of download time over different
populations in the same swarm.  (A different view of the same
experiment shown in Figure~\ref{figure:sample_efficiency}.)
\label{figure:sample_downloadtime}}
\end{figure}

Of course, when simulating a large population of nodes with various
configurations assigned at random, we would expect significant
variation in individual nodes' efficiency, even when they have the
same configuration.  Figure~\ref{figure:sample_efficiency} shows
cumulative distribution functions over nodes' efficiency in a simulation
with altruistic, standard, and leech nodes.  A curve that stays closer
to the bottom of the graph, as the altruistic data series does, represents
more nodes operating closer to their peak efficiency.
(This experiment shares the same configuration as used later in
Figure~\ref{figure:ignore_tyrants1}.)

While we could potentially generate a figure like this for every possible
simulation configuration, and every simulation run would generate a figure with the same general shape, this would obscure trends from one simulation to
the next.  Instead, we observe that the median value of each data series (i.e., the efficiency value for which the data series reaches 50\% on the $y$-axis)
represents an effective proxy for the overall behavior of the data.  If the
median values are close, then the curves will be close.  If the median values are far apart, then the curves will be far apart.

For our experiments, then, any given set of experimental parameters (as described in Section~\ref{sec:methodology:setup}) will yield three values: the median efficiency of each of the three populations (altruistic, standard, and leech), which we can then plot as we vary the simulation parameters.


An alternative to efficiency would be to consider the download times, without
normalizing them for differences in each node's available bandwidth.  
Figure~\ref{figure:sample_downloadtime} shows CDFs of download times
for the same experimental setup as Figure~\ref{figure:sample_efficiency}.
We added an ``optimal'' distribution, representing the best that the
altruistic nodes could ever have performed if they had achieved 100\% utilization of their download bandwidth.  We could have added additional ``optimal'' lines for each population, but this would make reading the figure more complicated.  
Furthermore, median values are less meaningful because the underlying distribution of bandwidths would vary if the random assignment were done differently.

Of course, absolute download time and download efficiency are measuring the same underlying phenomenon; improving one metric would clearly improve the other.
Table~\ref{table:median_compare} shows the median values from each of these figures.  The efficiency values elide unnecessary experimental details and concisely describe the relative performance of each population.

Lastly, we must convince ourselves that efficiency is a reliable metric
from one experimental run to the next.  Since many of the parameters in
our system are assigned randomly, we experimentally re-ran our
experiment twenty times, each time with a different random seed.
The results, shown in Table~\ref{table:exp1_stats}, show significant
variation from one run to the next, but the variations among altruistic
nodes are smaller than among standard nodes.
For an additional experiment, we changed the leech nodes to be
standard nodes.  We would expect, then, that they would behave the
same as standard nodes.  Table~\ref{table:exp2_stats} clearly
validates this behavior.

From these measurements, it appears that standard nodes are more
likely to be the victims of circumstance, while altruistic nodes and
leech nodes are more stable in the face of random variation.  As such,
the reported performance of standard nodes should be considered to be
noisier than the reported performance of altruistic or leech nodes.
While we could precisely work out the minimum change between different
populations that would represent a statistically significant
difference, this is insufficient for our needs.  Experimentally, we
must show that our desired altruistic behavior doesn't just make a
statistically significant improvement.  We must show a large enough
improvement to incentivize BitTorrent users to choose clients that
follow our desired behavior.

(For the remainder of the paper, we only run each experiment a single
time for a given set of experimental parameters.  Since each data point
takes as long as a day to compute, we cannot afford to run every experiment
twenty different times.)

%% file: evaluation.tex
In this section, we detail the findings of our research. We will first
demonstrate why seeding is important for swarms of nodes with asymmetric
bandwidth. We will then demonstrate how our algorithm improves performance
for seeding nodes. The next three subsections explore how bandwidth reservation,
altruistic population size, and rewarding node overlap impact the effectiveness
of our seeding algorithm. Finally, we analyze the performance of our algorithm
in swarms that include BitTyrant nodes.

\subsection{Importance of Seeding}

\begin{figure}
\centering
\includegraphics[scale=.3]{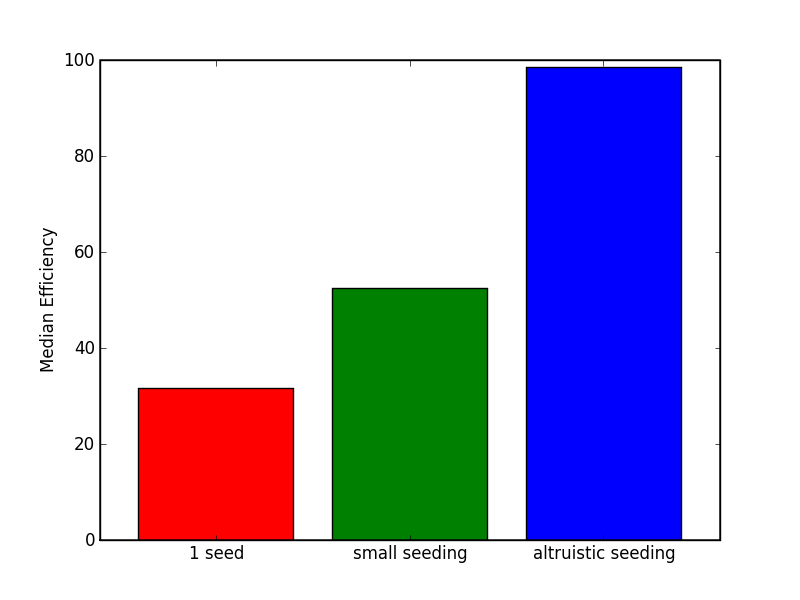}
\caption{The median efficiency of the overall swarm under different
compositions of clients. The worst performance is experienced when
there is only one seed. When 70\% of the clients 
seeding for 1-2 hours, the performance improves significantly. 
When 10\% of the nodes seed for 1-2 days, the median efficiency
approaches 100\%.
\label{figure:seeding_value}}
\end{figure}

Our first objective was to establish the importance of seeding to a BitTorrent
swarm. We ran our simulation with three different population configurations.
First, we ran the swarm with 1 initial seed and 100\% of the swarm composed 
of our leech clients that do no seeding whatsoever. Next, we ran the swarm with 
1 initial seed, 70\% of the standard clients that do a small amount of seeding,
and 30\% of the leech clients. Finally, we ran a simulation with 10\% altruistic
nodes that seed significantly, 70\% of the standard clients, and 20\% of the
leech clients. The results are shown in Figure~\ref{figure:seeding_value}.


There are two reasons why the swarm cannot obtain high efficiency without 
significant seeding contributions. First, the swarm is comprised of nodes with
asymmetric bandwidth profiles. In our swarm, the upload is always half of the
download capacity. Even with idealized operations, a swarm could hope for no more than
50\% efficiency from TFT trading alone.  The second issue is that a BitTorrent
swarm is not ideal. Various factors such as churn reduce the effectiveness of the
protocol.  Seeding provides enough additional capacity to overcome
these deficiencies.  

Clearly, seeding is essential for nodes in a swarm to maximize
their download bandwidth; if we can design a mechanism that
incentivizes more BitTorrent users to seed for longer periods,
this should have a clear, positive impact on the system.

\subsection{Rewarding Seeding}

To evaluate our reward seeding algorithm, we first ran a baseline simulation.
The setup for this simulation was 10\% altruistic
nodes, 70\% of the standard clients, and 20\% of the leech clients. All three
populations were running the standard BitTorrent trading and seeding algorithms, thus we expected 
all three populations to experience similar performance. As expected,
the results for all three populations was near 100\% efficiency.



\begin{figure}
\centering
\includegraphics[scale=.3]{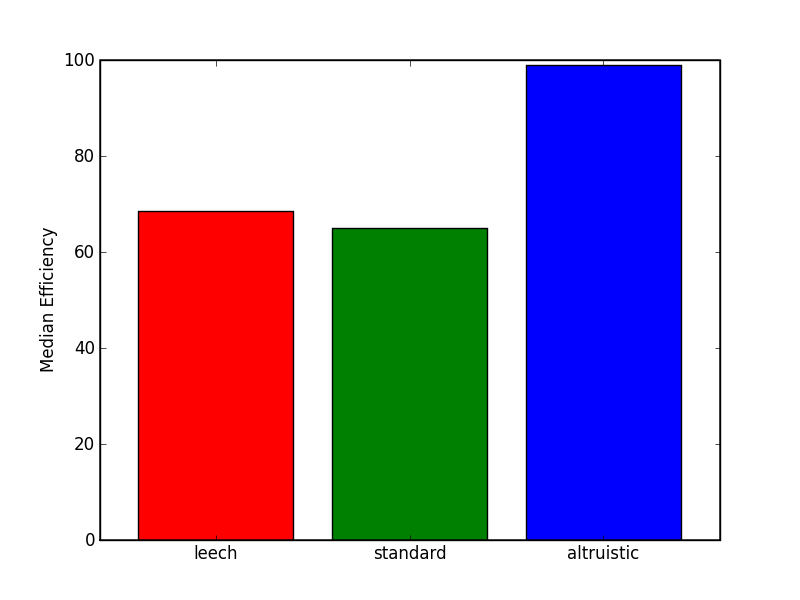} \\
\caption{Median efficiency 
when the altruistic population reserves 75\% of the seeding bandwidth for
other altruistic nodes.
\label{figure:baseline_rewarding}}
\end{figure}

We then repeated this baseline experiment with all of the altruistic nodes configured 
to run our reward seeding algorithm, reserving 75\% of their bandwidth
for rewards to prior seeders.
The other two populations continued to use normal seeding algorithms. 
In this version of our experiment, we assumed
perfect overlap for this altruistic group. In other words, 
every altruistic node had been previously seeded by every other
altruistic node, prior to the start of the experiment, and would
thus allow the other altruistic nodes to share in the bandwidth
reserved for rewards.
The results of this simulation are shown in Figure~\ref{figure:baseline_rewarding}. 
The altruistic population maintained
nearly perfect efficiency, while the two unrewarded populations experienced a 
significant drop in performance.


\subsection{Bandwidth Reservation}
\label{section:reservation}

\begin{figure}
\vskip -1em
\centering
\includegraphics[scale=.4]{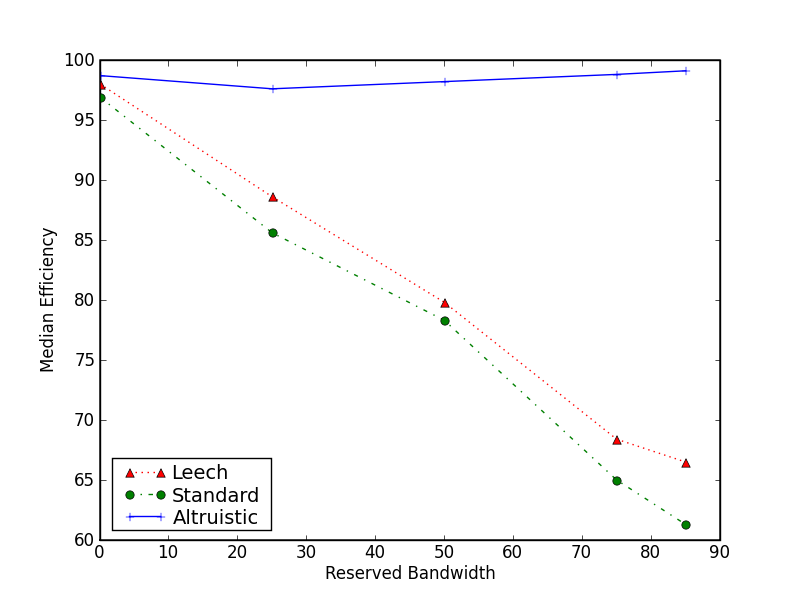}
\caption{Median efficiency 
as a function of the reserved bandwidth by the altruistic nodes.
\label{figure:reserved_bandwidth}}
\centering
\includegraphics[scale=.4]{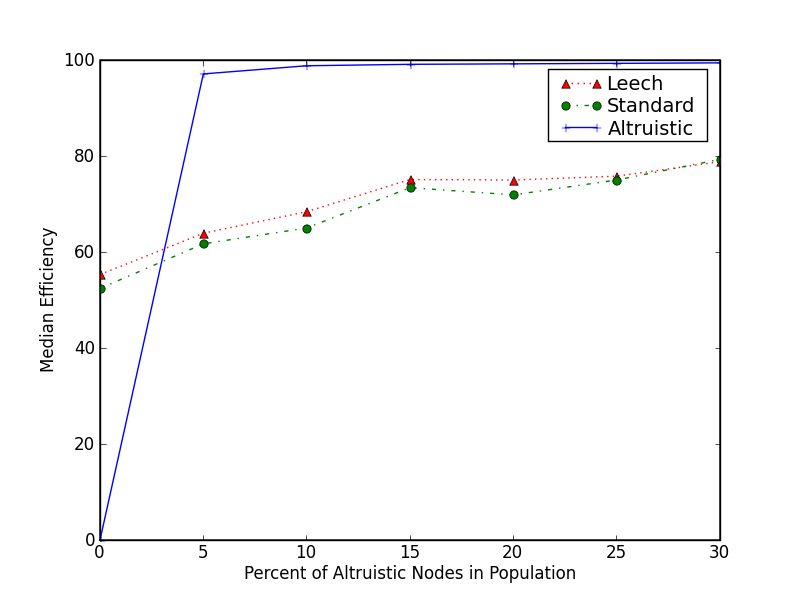}
\caption{Median efficiency 
as a function of the percentage of altruistic nodes in the swarm.
\label{figure:altruistic_size}}
\centering
\includegraphics[scale=.4]{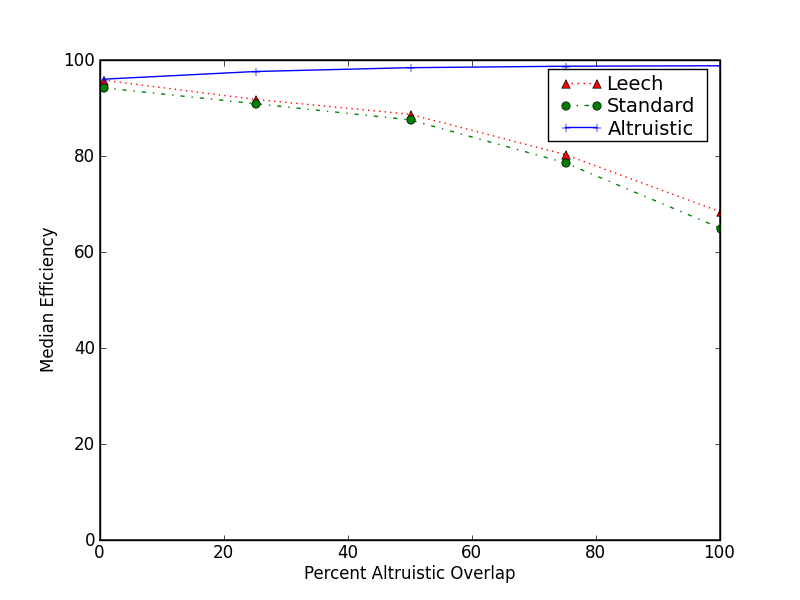}
\caption{Median efficiency 
as a function of the percentage of overlap in the altruistic nodes.
\label{figure:overlap}}
\end{figure}

As described before, our seeding algorithm can reserve bandwidth
for the exclusive use of nodes being rewarded. 
To understand the necessity of these bandwidth reservations,
we ran a simulation where we varied the percentage of reserved
vs. unreserved seeding bandwidth.  The results, shown in 
Figure~\ref{figure:reserved_bandwidth}, show
the median efficiency of the altruistic,
standard, and leech populations in simulations with different reserved bandwidth
configurations. In all simulations, there are 10\% altruistic, 70\% standard, and 20\%
leech clients.  The bandwidth reservation applies to altruistic nodes'
seeding bandwidth.  For the moment, we are assuming that altruistic nodes
all have prior history and know which other nodes have seeded in the
past.

One immediate observation is that 
our seeding algorithm, without any bandwidth reservation, does no better than
normal seeding. This seems counter-intuitive because
the rewarded nodes should still be getting more seeded bytes than their unrewarded
peers. One might think that there would be some performance
improvement for the altruistic nodes, even
with 0\% reserved bandwidth, but they are already getting nearly 100\%
efficiency.

With bandwidth reservations,
if there is insufficient demand
from the ``reward'' population, then that portion of the seeding bandwidth 
will go unused.  In short, our work suggests that the only way to
create a performance differential between rewarded and non-rewarded
nodes
is to withhold service from unrewarded nodes.


There is an interesting trade-off, however.  If the reservation is too
high, then all of the bandwidth is effectively being spent on
maintaining old relationships rather than establishing new ones.
As nodes quit old swarms and join new ones on a regular basis,
there is a clear incentive to have seeded to strangers in the past
if there might be a payout in the future.


\subsection{Altruistic Population Size}
\label{section:altruistic_size}

We cannot predict what percentage of nodes in a given swarm might be running our reward seeding algorithm.
We would like to verify, regardless of the breakdown, that incremental growth in the 
reward seeding group will yield benefits both for those nodes as well as for everybody else.
This leads to the question of how the system will respond as the
population dynamics change.
Figure~\ref{figure:altruistic_size} shows how efficiency changes
as a function of the percentage of the altruistic and standard populations in the
total swarm. The leech population is fixed at 20\% and 
the rewarding nodes reserve 75\% of their bandwidth.

This experiment demonstrates that the performance of the entire swarm improves as more nodes follow our altruistic scheme, even when reserving 75\% of their bandwidth for reward seeding.  That other 25\% is enough to improve things for everybody else.

At some point, beyond the 30\% altruism rate where we terminated our simulation, the standard nodes may have sufficient efficiency that they would be disincentivized to change to the altruism strategy.  By then, the altruism strategy would already be the dominant behavior in the swarm.  Also, regardless of the rate of altruistic nodes, this experiment shows that altruism {\em always} wins, and sometimes wins big, even with relatively low populations of altruistic nodes.

\subsection{Overlap}
In this section, we explore the highly critical overlap parameter. Our algorithm
assumes that nodes are rewarding based on first-hand information gleaned from prior interactions in prior swarms. In 
previous experiments, we have assumed that this knowledge of prior interactions, which we call {\em overlap}, is complete.  Every node has prior, positive interactions with its altruistic peers and thus knows to include them in the reward population during future interactions.
Such oracular knowledge is not realistic.

For simulation purposes, we wish to vary the degree to which
altruistic nodes have had past interactions with other altruistic
nodes and thus have the first-hand knowledge necessary to give reward
seeding to their peers.  To accomplish this, we partition the
altruistic nodes into two sub-groups: rewarding and non-rewarding
nodes. Rewarding nodes will reward all other altruistic nodes,
including non-rewarders, while non-rewarding nodes will reward nobody.
Non-rewarding nodes still have the same 75\% bandwidth reservation, but
they never use it.  By varying the ratio of rewarding to non-rewarding nodes,
we can roughly simulate the real-world effects that might be seen as the
degree of overlap between altruistic nodes varies.

Figure~\ref{figure:overlap} shows the efficiency for each population as a function of the percentage of altruistic nodes that are rewarders.
We maintain a 10\% altruistic, 70\% standard, and 
20\% leech population. Reserved bandwidth remains fixed at 75\%.

Our experiment demonstrates that overlap is clearly necessary to achieve the benefits of our reward seeding strategy.  Once the overlap reaches 50\% (i.e., about half of the seeding interactions between altruistic nodes are rewarded with higher bandwidth), the performance improvement for the altruistic strategy is undeniable.  Whether such an overlap rate can be achieved in the real world is unclear.  We discuss some strategies that might compensate for this in Section~\ref{discussion}.

\subsection{Seeding Rewards versus BitTyrant}
\label{section:eval-bittyrant}

In this section, we test the altruistic reward seeding algorithm against clients running
the more aggressive BitTyrant trading algorithm. BitTyrant clients tend to see
improved performance at the expense of other nodes in the system.  (BitTyrant was introduced in Section~\ref{sec:background:ambient}.)

\begin{figure}
\vskip -2em
\centering
\includegraphics[scale=.4]{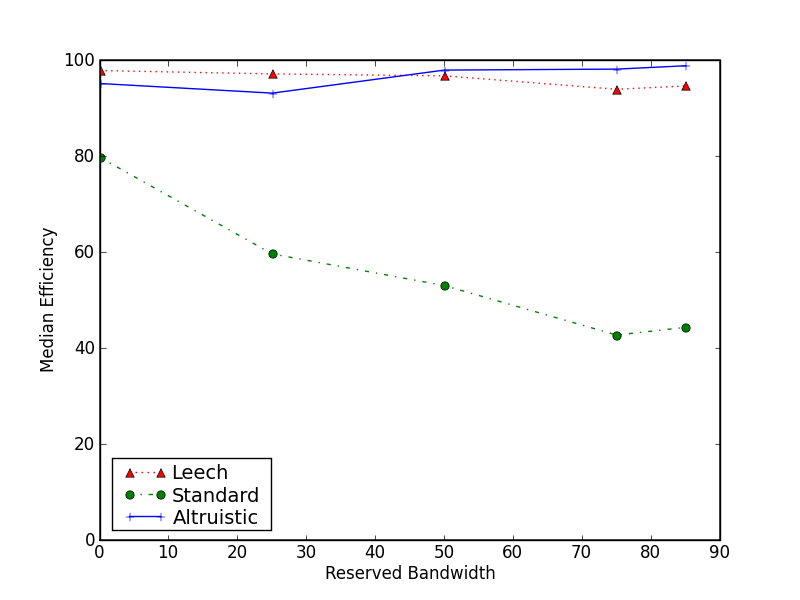}
\caption{Altruistic nodes versus tyrants under different amounts of
reserved bandwidth.
\label{figure:normal_v_tyrant_reserved}}
\centering
\includegraphics[scale=.4]{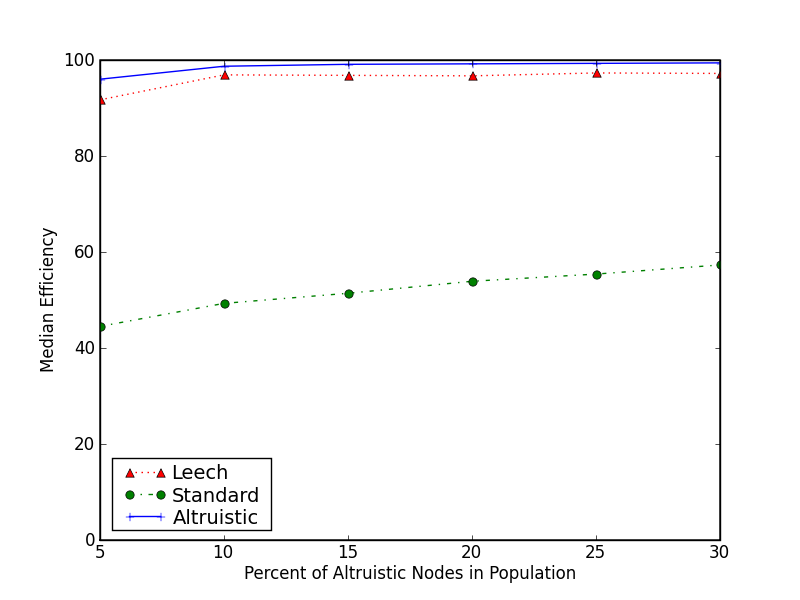}
\caption{Altruistic nodes versus tyrants with different ratios of
altruistic nodes in the population.
\label{figure:tyrant_altruistic_size}}
\centering
\includegraphics[scale=.4]{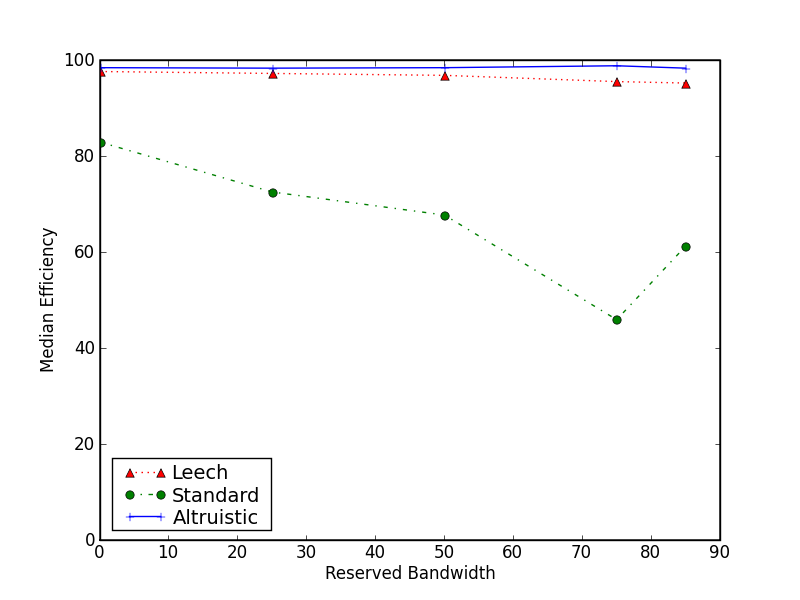}
\caption{Reward-seeding altruists, modified to trade tyrannically before they begin seeding, versus tyrant-leeches under
different amounts of reserved bandwidth.
\label{figure:tyrant_reserved}}
\end{figure}

Our first experiment, shown in Figure~\ref{figure:normal_v_tyrant_reserved}, pits rewarding seeders against tyrannical leeches.
This test repeats the
bandwidth reservation experiment of Section~\ref{section:reservation}
with the leeching population configured to use the BitTyrant trading
algorithm. All other parameters remain the same.

Comparing these results against those of the earlier bandwidth reservation
test, we note that BitTyrant-leeches performed as well as the rewarded
altruists. At the same time the leeches degraded the performance of the
standard nodes significantly. From this we conclude that the reward-seeding 
algorithm protects against, or at least ameliorates the exploitation of the BitTyrant protocol,
but that it does not sufficiently penalize the leeching clients.

To evaluate how the size of the altruistic population impacts the
performance of these populations, we repeated the experiment of
Section~\ref{section:altruistic_size}, again with the rewarding
altruistic seeders versus the tyrannical leeches. We hoped that
increasing numbers of altruists might be able to penalize the
tyrannical leeches. Unfortunately, as shown in
Figure~\ref{figure:tyrant_altruistic_size}, the tyrannical leeches
still had no trouble achieving near perfect efficiency.

We considered the possibility that the leeching nodes would
not do so well if the altruistic nodes were more stingy during
the TFT trading phase. To test this, we reconfigured
the bandwidth reservation test.  In this experiment, the altruists
use the BitTyrant TFT strategy rather than the default
BitTorrent TFT strategy, but still perform the incentivized reward seeding.
The leech population still practices tyrannical TFT trading and never seeds.
The standard population uses
standard algorithms for both seeding and TFT trading. All other simulation 
parameters remained the same. The results are shown in Figure~\ref{figure:tyrant_reserved}.

Based on these experiments, a rational
user might just as well run a tyrannical client as an altruistic client.
They will receive the same download efficiency and they will minimize their
upload bandwidth.

\subsection{BitTyrant Exploitation}
\label{sec:tyrant-exploit}
In the pursuit of finding a weakness in BitTyrant's seemingly anti-social
behavior, we
discovered a problem with BitTyrant's exchange mechanism (also noted by Carra et al.~\cite{tyrant_bad_eval}). The original BitTyrant
paper~\cite{bittyrant1_2007} says:

\begin{quote}
As such, BitTyrant continually reduces send rates for peers that reciprocate,
attempting to find the minimum rate required. Rather than attempting to ramp up 
send rates between high capacity peers, BitTyrant tends to spread available capacity
among many low capacity peers, potentially causing inefficiency due to TCP
effects.

To work around this ... effect, BitTyrant advertises itself at connection time using
the Peer ID hash. Without protocol modification, BitTyrant peers recognize one another and
switch to a block-based TFT strategy that ramps up send rates until
capacity is reached. 
\end{quote}

\noindent
The authors believe that their weakness is looking for too
many low bandwidth flows, or that the many low bandwidth flows are inefficient because 
of TCP effects.

To evaluate this, we ran several simulations without the BitTyrant block-level TFT component (i.e., we disabled BitTyrant's ability to detect that a peer is also running BitTyrant).
BitTyrant nodes did very poorly when communicating with
each other.

BitTyrant assumes it is receiving reciprocation when it receives an unchoke. This is
a valid assumption for BitTorrent nodes, but it is not as clear of a signal from another
BitTyrant node because it does not indicate how much they are willing to upload. So,
if two BitTyrant nodes unchoke each other, they both assume they have an estimate for the minimum upload speed necessary to achieve reciprocation.  They then  both begin
to drop their upload rates potentially down to zero in a quest to achieve lower estimates for the minimum upload speed.


\begin{figure}
\centering
\includegraphics[scale=.3]{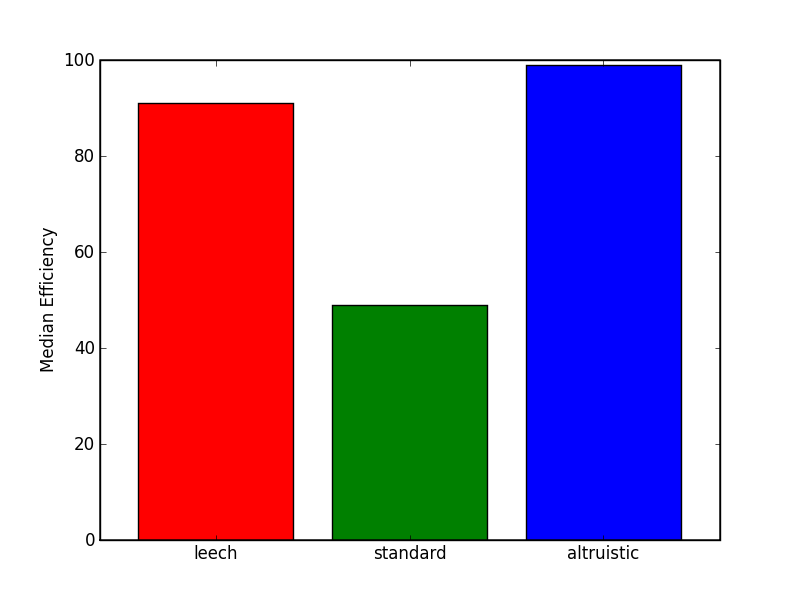}
\caption{Median efficiency when altruistic nodes refuse to seed anything to tyrannical leech nodes.
\label{figure:ignore_tyrants1}}
\end{figure}

BitTyrant solves this problem by self-identification, disabling the reciprocation-discovery mechanism because it doesn't really work between two tyrants.
This identification features can be exploited by
altruistic nodes to deny service to tyrants!
A BitTyrant node cannot lie or obscure that it's a tyrant
without incurring a penalty when trading with other tyrants.

We re-ran our baseline simulation with 10\% altruistic, 70\% standard, and 20\% leech nodes.
The altruistic nodes used the normal trade algorithm and our reward seeding algorithm.
The leech nodes used the BitTyrant trade algorithm. Bandwidth was reserved at 75\%
and the altruistic nodes ignored tyrants during seeding, but interacted with them normally when still downloading the torrent. The results are shown
in Figure~\ref{figure:ignore_tyrants1}.

By ignoring tyrants, the altruistic nodes achieve a small but
significant performance improvement relative to the tyrants. There may well be other ways to exploit
tyrants, such as refusing to interact with them at all.
It is sufficient to say that BitTyrant is
vulnerable to exploitation, itself, as a consequence of its necessary self-identification mechanism.

%% file: discussion.tex
The development of this research gives rise to a number of
important discussion points that we will address here. These
points include issues relating to the practicality of our algorithm
to real-life solutions as well as topics of future research.

\paragraph{Privacy / Anonymity} is of significant concern for many BitTorrent 
users. Naturally, 
a long-term identifier would impact anonymity. However, the
BitTorrent protocol was never engineered to provide anonymity to
BitTorrent users.  (They announce their presence to everybody in the
swarm, based on their IP address,
and advertise what pieces they have available to trade!)
From this perspective, a long-term identifier is not much worse
than an IP address.

On the other hand, if a BitTorrent user chose to tunnel
BitTorrent through an anonymization system like Tor, then the
IP address would be obscured, while the long-term identifier would
still be advertised.  While a number of BitTorrent users do tunnel
traffic through Tor, their performance will suffer greatly, as Tor was
never intended to support the kind of massive, sustained traffic flows
that BitTorrent can generate.  Engineering an anonymity service
specifically for BitTorrent would be an interesting opportunity for
future research.

\paragraph{Bootstrapping and Overlap} are the most critical concerns for further
development of this incentives mechanism. The reward mechanisms in
our research depend on the same nodes seeing one another, again and
again.  This may not occur much, in the general case, but it could
well happen in particular subcommunities.

{\em Existing Small Groups}\/: A number of relatively small
(compared to the entire world) communities exist for the purpose
of BitTorrent distribution. The traces we described in
Section~\ref{methodology} were collected from \url{filelist.org} over a three month period. This
community requires a sign-in name which was associated with
each download. We observed that 50\% of all peers participated 
in at least two of the same swarms. These types of groups 
would be able to switch over to the seed-rewarding algorithm with
very little difficulty and would likely have sufficient overlap.

{\em Social Groups}\/: Existing social communities, brought together
by mutual interests on social networks, could be used to leverage a 
relatively small BitTorrent community suitable for the seed-rewarding
algorithm.

{\em Shared Interests}\/: Even without explicit social groupings, one
would reasonably expect that many people will follow similar
patterns. For example, a variety of television shows are distributed
via BitTorrent.  Users who download the current show are likely to
download subsequent shows.  Similar affinities would be expected
around other content that is updated on a regular basis, such as
updated Linux distributions.

{\em Transitive Trading} and similar methods, may be able to
ameliorate the need for extensive overlap. Transitive
trading~\cite{NNS+fudico04,scrivener05}
allows two clients that have never met to exchange ``credits'' through a mutual
contact.

\paragraph{BitTyrant} is an important development in BitTorrent because
it improves the efficiency of certain core concepts. For example, the
optimistic unchoke in standard BitTorrent trading is a \textit{search} method
for finding better peers, but it simply searches randomly.
However, as we discussed in Section~\ref{sec:tyrant-exploit}, BitTyrant clients must identify whether
they are speaking to other tyrants and change strategies.  Otherwise, the
default BitTyrant TFT strategy will have both clients dropping their
bandwidth all the way to zero.  

This BitTyrant flaw creates interesting opportunities.
Since BitTyrant
clients must identify themselves as such, they can be trivially
ignored by other clients who, perhaps, do not with to support
their tyrannical behavior.  However, there are many other options.
BitTyrant clients (or, really, any BitTorrent client) could publish
categorical statements about their unchoking policies.  For example
a node might declare: ``If you give me
at least $X$ bytes per second, then I'll unchoke you and give you 
$X$ in return, up to $Y$ bytes per second max.''  Of course, a tyrant
could lie about such policies, but it creates an interesting
opportunity for future research, both in terms of simulation studies and
in terms of economic modeling.

Carra et al.~\cite{tyrant_bad_eval} also examined the strengths of BitTyrant-style 
behavior versus simply expanding the number of simultaneous connections in 
traditional BitTorrent clients by simulation. However, their simulation
models ignored churn and other real-world conditions leading us to believe
that their results are unreliable.


%% file: related.tex
The BitTorrent protocol and associated algorithms were introduced by
Cohen in 2003~\cite{bittorrent-incentives} with a reference client
implementation. A fluid model for the system was given by Qiu et
al.~\cite{bt_performance_2004}, who used it to show that in certain
cases a Nash equilibrium can exist in systems where peers
choose upload rates to match their download rates. Studies performed
on emulated swarms by Legout et al.~\cite{liogkas} validated the
effectiveness of the BitTorrent unchoking algorithm. Legout et
al.~\cite{legout2} also concluded from real-world tests that the
rarest-first algorithm is very important to system performance, and
argued that the default unchoking algorithm provides adequate robustness
from free-riders.

A fluid-model simulator was used by Bharambe et
al.~\cite{msr_simulator_2005} to represent a BitTorrent system in a
more abstract manner than our own. They confirmed the utility of the
rarest-first policy for piece selection. They also investigated
unfairness with respect to volume uploaded and argued that the
rate-based TFT strategy fails to prevent such unfairness, especially
in systems with a great disparity of bandwidth among peers. They
proposed a new block-level, volume-based TFT trading algorithm,
although subsequent researchers challenged its
effectiveness~\cite{legout2}.

De Vogeleer et al.~\cite{vogeleer}, made an event-based simulator for
BitTorrent based on the algorithms in the reference implementation and
used it to model a variety of typical swarm scenarios, verifying the
performance characteristics against the expected behavior of a
standard BitTorrent client.

A simulation-based study by Eger et al.~\cite{egeretal_upgrade_07}
compared flow-level and packet-level simulations for BitTorrent-like
systems and found that, while flow-level simulations are useful for
demonstrating the theoretic performance of the de facto BitTorrent
scheme, the delay of TCP packets and other cross-layer effects have a
significant impact on BitTorrent performance, and these effects
require a more granular simulation to be adequately captured.

Much research has been performed concerning the robustness of
BitTorrent's tit-for-tat trading mechanism against selfish
behaviors. BitTorrent was modeled as a form of the Iterated Prisoner's
Dilemma problem by Jun et al.~\cite{jun}, who suggested that the
current peer-selection algorithm is susceptible to free-riders; they
proposed a different TFT strategy. Tian et al.~\cite{tian} used
mathematical models as well as simulation-based and real-world
experiments to argue for a modified TFT algorithm.

Sirivianos et al.~\cite{sirivianos} emulated a strictly free-riding
client which contacts the tracker often to gain a large neighborhood
from which to free-ride; they concluded that this attack was feasible
in practice.
Liogkas et al.~\cite{liogkas} use PlanetLab to demonstrate three
different exploits: downloading from seeds, downloading from the
fastest peers, and advertising fake pieces.


%% file: conclusion.tex
At present, BitTorrent's seeding mechanism
is entirely altruistic; nodes have no rational reason to offer
seeding service to their peers, yet the additional bandwidth provided
by seeding is essential to the efficient operation of BitTorrent.
Anything that can encourage seeding would have an immediate knock-on benefit for
BitTorrent users.

In this work, we have proposed a method for rewarding seeding in
BitTorrent by means of long-term identification. 
Nodes remember peers that
seeded to them in the past and reciprocate by seeding to them
in later swarms.

To evaluate our algorithm and its parameter space, we developed and
employed a flow-level simulator. The algorithm was tested on realistic
file-sizes and trace-driven churn to improve its accuracy. We found
that our algorithm improved the download efficiency of the BitTorrent
nodes from 70\% to 95\% or better. This improvement represents the
upper bound of our algorithm's performance and was based on oracular
knowledge that would not be available in real scenarios. We tested
more realistic settings and found that our algorithm could still
increase the download efficiency by ten percentage points.

Finally, we evaluated our seed-rewarding algorithm in swarms that
had some portion of the population running BitTyrant, a variant
on BitTorrent that is aggressive about getting fast downloads
with minimal investments of upload bandwidth.
We found that our algorithm could protect nodes from being exploited
by BitTyrant, but
could not sufficiently penalize tyrannical behavior to discourage users
from choosing to run BitTyrant.
However, leveraging a weakness in BitTyrant, where BitTyrant nodes
must identify themselves as such,
we can ignore tyrants during seeding
and reduce their performance.

So long as BitTorrent peers have sufficient overlap in successive
swarms, allowing them to build individual long-term histories of who has
seeded in the past, we conclude that BitTorrent peers using our incentivized reward
seeding algorithm will enjoy better performance for themselves
and also improve performance for their peers, whether running
our algorithm or not.
By adding in our mechanism, for which peers have a genuine incentive to follow, we can build better robustness in BitTorrent.